\documentclass[draft]{agujournal2019}
\usepackage{url} 
\usepackage{lineno}
\usepackage{amssymb,amsfonts,amsmath}
\usepackage{wasysym,gensymb,bm} 
\usepackage[inline]{trackchanges} 
\usepackage{mathtools}
\usepackage[normalem]{ulem}
\draftfalse
\journalname{Geophysical Research Letters}
\begin{document}
\title{Cloud Feedback on Earth's Long-term Climate Simulated by a Near-global Cloud-permitting Model}
\authors{Mingyu Yan\affil{1}, Jun Yang\affil{1}, Yixiao Zhang\affil{1,2}, Han Huang\affil{1,3}}

\affiliation{1}{Laboratory for Climate and Ocean-Atmosphere Studies, Department of Atmospheric and Oceanic Sciences, School of Physics, Peking University, Beijing 100871, China}
\affiliation{2}{Now at Department of Earth Atmospheric and Planetary Sciences, Massachusetts Institute of Technology, 77 Massachusetts Avenue, Cambridge, MA 02139, USA}
\affiliation{3}{Now at Department of Atmospheric and Oceanic Sciences, McGill University, Montreal, Canada}
\correspondingauthor{Jun Yang}{junyang@pku.edu.cn}

\begin{keypoints}
\justifying  
\item A near-global cloud-permitting model with a resolution of 10~km\,$\times$\,14~km is employed to test whether there is a long-term cloud feedback.
\item The cloud feedback does have a net cooling effect when the Sun becomes brighter and meanwhile the CO$_2$ concentration decreases.
\item These results confirm that cloud feedback is a part of the solution to the faint young Sun problem but its magnitude is relatively small.
\end{keypoints}

\begin{abstract}
\justifying
The Sun becomes brighter with time, but Earth's climate is roughly temperate for life during its long-term history; for early Earth, this is known as the Faint Young Sun Problem (FYSP). Besides the carbonate-silicate feedback, recent researches suggest that a long-term cloud feedback may partially solve the FYSP. However, the general circulation models they used cannot resolve convection and clouds explicitly. This study re-investigates the clouds using a near-global cloud-permitting model without cumulus convection parameterization. Our results confirm that a stabilizing shortwave cloud feedback does exist, and its magnitude is $\approx$6~W\,m$^{-2}$ or 14\% of the energy required to offset a 20\% fainter Sun than today, or $\approx$10~W\,m$^{-2}$ or 16\% for a 30\% fainter Sun. When insolation increases and meanwhile CO$_2$ concentration decreases, low-level clouds increase, acting to stabilize the climate by raising planetary albedo, and vice versa.
\end{abstract}

\section*{Plain Language Summary}
\justifying
The emergence and evolution of life require a relatively stable climate environment. In the solar system, life has been found only on Earth and appeared since about four billion years ago. The underlying mechanisms for maintaining the long-term climate on Earth is an important question but the answer is not completely clear. In this study, we re-investigate a recently proposed mechanism, a stabilizing cloud feedback, using a high-resolution cloud-permitting model in a near-global domain. Our simulations confirm that a stabilizing cloud feedback does exist, but its magnitude is relatively small.

\section{Introduction}
\justifying   
The luminosity of the Sun increases with time, and the solar constant in the Archean Eon was 20--30$\%$ lower than that today \cite{gough1981}. If other climate-controlling factors were the same as present, Earth would have been in a globally ice-covered snowball state during the Archean \cite{sagan1972}, but much evidence indicates that there was surface liquid water \cite{feulner2012}. Hence, there should be some factors compensating for this discrepancy \cite{sagan1972,feulner2012,charnay2020}, and it is one of the most important problems on Earth's long-term climate \cite{pierrehumbert2010}. A number of studies show that biogeochemical feedbacks, such as the carbonate-silicate weathering cycle, are key to stabilizing the climate. The increase of greenhouse gas contents, such as CO$_2$ and CH$_4$, is likely responsible for the FYSP (e.g., \citeA{wolf2013, byrne2014, le2014}). Other factors or feedbacks may have also contributed to solving this problem, such as a different atmospheric pressure, a less continent coverage, N$_2$-H$_2$ collision-induced warming, and a stronger tidal heating induced by the closer Moon \cite{goldblatt2009,rosing2010,goldblatt2011, wordsworth2013,heller2021}. A recent study of \citeA{Goldblatt2021} suggested that a long-term cloud feedback exists and it could also be a part of the solution to the FYSP.

Clouds are one of the key factors that determine planetary climate. For low-level clouds, the cooling effect dominates; but for high-level clouds, the greenhouse warming effect dominates \cite{hartmann2005, pierrehumbert2010}. \citeA{goldblatt2011} explored how clouds could resolve the FYSP in one-dimensional (1D) radiative-transfer simulations. They showed that a maximum radiative effect change of 15~W\,m$^-{}^2$ can be traced to a plausible low-level cloud reduction or high-level cloud increase. Using 3D GCMs, \citeA{charnay2013} and \citeA{wolf2013} focused on the Archean Eon, and they found that the changes of clouds have a warming effect on the surface, mainly due to the reduction of low-level clouds. \citeA{Goldblatt2021} systematically studied the cloud feedback using two GCMs: CAM4 and CAM5. They kept almost the same global-mean surface temperature ($|\Delta T| \textless 0.5 K$) by increasing solar constant and meanwhile decreasing CO$_2$ concentration. Their work showed that shortwave cloud feedback has a stabilizing effect on the long-term climate, and it can offset 20~W\,m$^-{}^2$ or 40\% in percentage of the required forcing when the insolation decreases from 1.0 to 0.8 times modern value. Below, we will show that they have overestimated the strength of the shortwave cloud feedback because the direct effect of varying solar radiation on the magnitude of shortwave cloud radiative effect without any change of cloud properties was not excluded. After subtracting this part, the magnitude of the shortwave cloud feedback in CAM4 is 8.6 W\,m$^-{}^2$ or 18.5\% (see Section~3.1).

The horizontal grid sizes of GCMs are always larger than $\approx$100~km, but the sizes of convection and clouds are in the order of $\mathcal{O}$(1--10)~km or even less, so that empirical convection and cloud parameterizations are necessary for GCMs. However, for GCMs, cloud feedback is the largest uncertainty source, leading to significant inter-model differences (e.g., \citeA{soden2006,vial2017}). Therefore, it is important to use higher-resolution models or models with different parameterization schemes to re-examine the long-term cloud feedback proposed in \citeA{Goldblatt2021}. One method is using a cloud-resolving model that has a resolution of $\mathcal{O}$(1)~km or smaller, but unfortunately, global-scale cloud-resolving simulations are far beyond present computation resources, especially when long-time integration and multiple experiments are required. Here, we use a cloud-permitting version with a resolution of $\approx$10~km, which is coarser than cloud-resolving models but much finer than GCMs, to re-simulate the clouds and the cloud feedback.

\section{Model Descriptions and Experimental Designs}

\begin{figure}
\noindent\includegraphics[width=\textwidth]{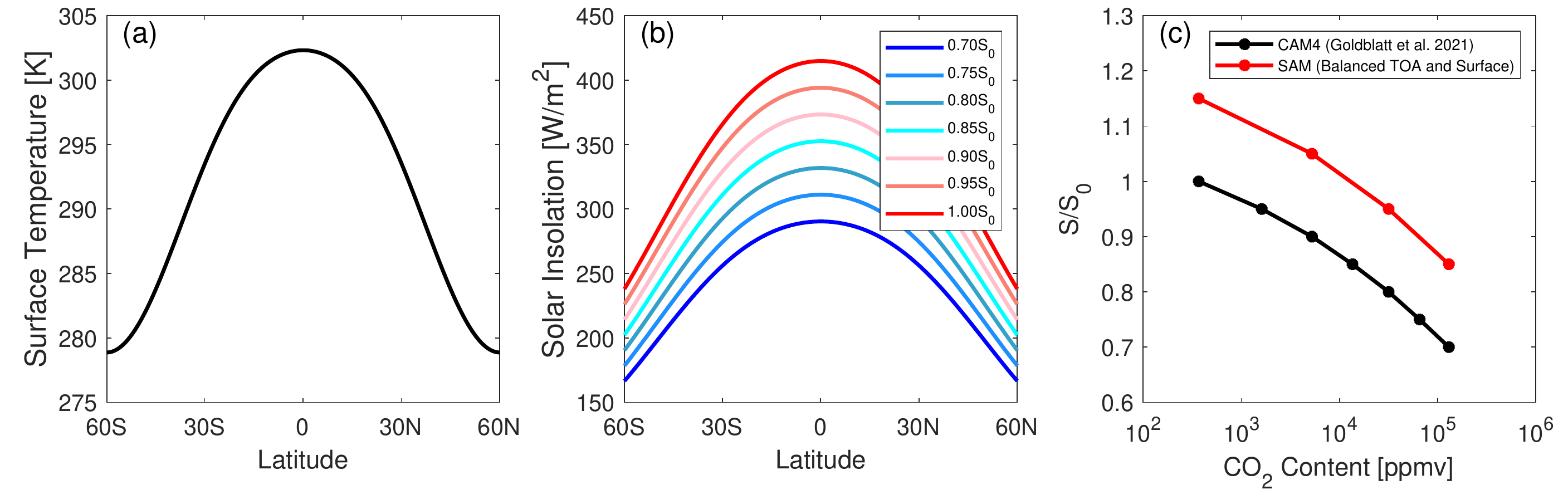}
\caption{(a) Zonal-mean sea surface temperature in fixed SST simulations, (b) solar insolation distribution, and (c) CO$_2$ mixing ratio and insolation in different simulations.}
\label{experimental}
\end{figure}

We run a series of simulations with the version 6.11.6 of the System for Atmosphere Modeling (SAM, \citeA{Khairoutdinov2003}). Anelastic momentum equations are used to explicitly resolve non-hydrostatic flows instead of the hydrostatic approximation used in GCMs. Radiation scheme is the same as the  Community Atmosphere Model version 3 (CAM3) \cite{collins2006}, and it is the same as that used in the CAM4 simulations of \citeA{Goldblatt2021}. There are still parametrizations in SAM that are used to calculate microphysics and sub-grid fluxes. Microphysics parameterization is based on a simple one-moment scheme, and a Smagorinsky-type closure scheme is used to calculate subgrid-scale fluxes \cite{khairoutdinov1999,Khairoutdinov2003}.

We use an aquaplanet mode, which means there is no land, and the northern and southern hemispheres are symmetric. The horizontal resolution is $\approx$14 km in latitude and $\approx$10 km in longitude, and the domain size corresponds to 60$^{\circ}$S to 60$^{\circ}$N in latitude and 0$^{\circ}$ to 180$^{\circ}$ in longitude. Rigid boundary conditions are used at $y$ boundaries, and periodic boundary conditions are used at $x$ boundaries. There are 38 vertical levels up to $\approx$35~km. The lowest mid-layer level is at 35~m, and the vertical grid spacing is $\approx$80~m between the two lowest levels and gradually increases to 1500~m above 10~km. A sponge layer is included above 28~km to reduce the reflection of gravity waves near the model top.

Letting sea surface temperature (SST) be interactive in a near-global domain comes with a huge computational cost, so we prescribe the SST. The spatial pattern of the specified SST is based on modern Earth, but it is zonally uniform and hemispherically symmetric (Figure~\ref{experimental}a). We neglect diurnal and seasonal cycles, and set the distribution of solar insolation to be zonally uniform but latitude dependent (Figure~\ref{experimental}b). For simplicity, the concentrations of ozone are set to 10$^-{}^7$ g g$^-{}^1$ for all layers, and there is no other trace gas. Besides, cartesian geometry is used in SAM.

We did seven experiments by changing CO$_2$ concentration and solar constant synchronously (black line in Figure~\ref{experimental}c). These experiments are respectively labeled as 0.70S$_0$, 0.75S$_0$, 0.80S$_0$, 0.85S$_0$, 0.90S$_0$, 0.95S$_0$, and 1.00S$_0$. The choices of insolation and CO$_2$ concentration are based on \citeA{Goldblatt2021}, in which a coupled slab ocean was employed and the obtained global-mean surface temperatures are nearly the same although the equator-to-pole surface temperature gradients change somewhat. For each experiment, the last 30 days are used in the following analyses (Figure~S1). The details of these simulations are shown in Text S1.

Hourly-mean cloud fraction, precipitation, and precipitable water in the modern Earth simulation (1.00S$_0$) show that the key characteristics, such as the Intertropical Convergence Zone (ITCZ) and mid-latitude baroclinic clouds, can be well simulated in the cloud-permitting framework (Figure~S2). However, there are some less realistic features, which were also noted in \citeA{bretherton2015b}. For example, the differences between continental clouds and oceanic clouds are not simulated because of the employed aquaplanet mode, and neither seasonal cycle nor diurnal cycle is simulated because of the fixed solar radiation and SST. Moreover, as we set the boundaries at 60$^\circ$S(N), a lot of clouds are trapped near the boundary walls; these clouds are mostly advected from lower latitudes and trapped by the walls; if there were not these boundaries, clouds will further flow to higher latitudes and then precipitate to the surface.

Note that energy budgets at the top of the atmosphere (TOA) and the surface are not balanced because we prescribe SSTs. To decrease the energy imbalances of the TOA and surface, we re-do four experiments (0.70S$_0$, 0.80S$_0$, 0.90S$_0$, and 1.00S$_0$) with the solar constants increasing to respectively 0.85, 0.95, 1.05, and 1.15~S$_0$, and meanwhile the CO$_2$ concentrations are unchanged (Figure~S3 and red line in Figure~\ref{experimental}c). Besides, we also add two slab ocean runs (see Text S1, Figure S4).  In these experiments, the energy imbalances at the TOA and surface decrease to be less than 3 W\,m$^-{}^2$ (Figures~S3 and S4).


\section{Results}

\begin{figure}[t]
\noindent\includegraphics[width=\textwidth]{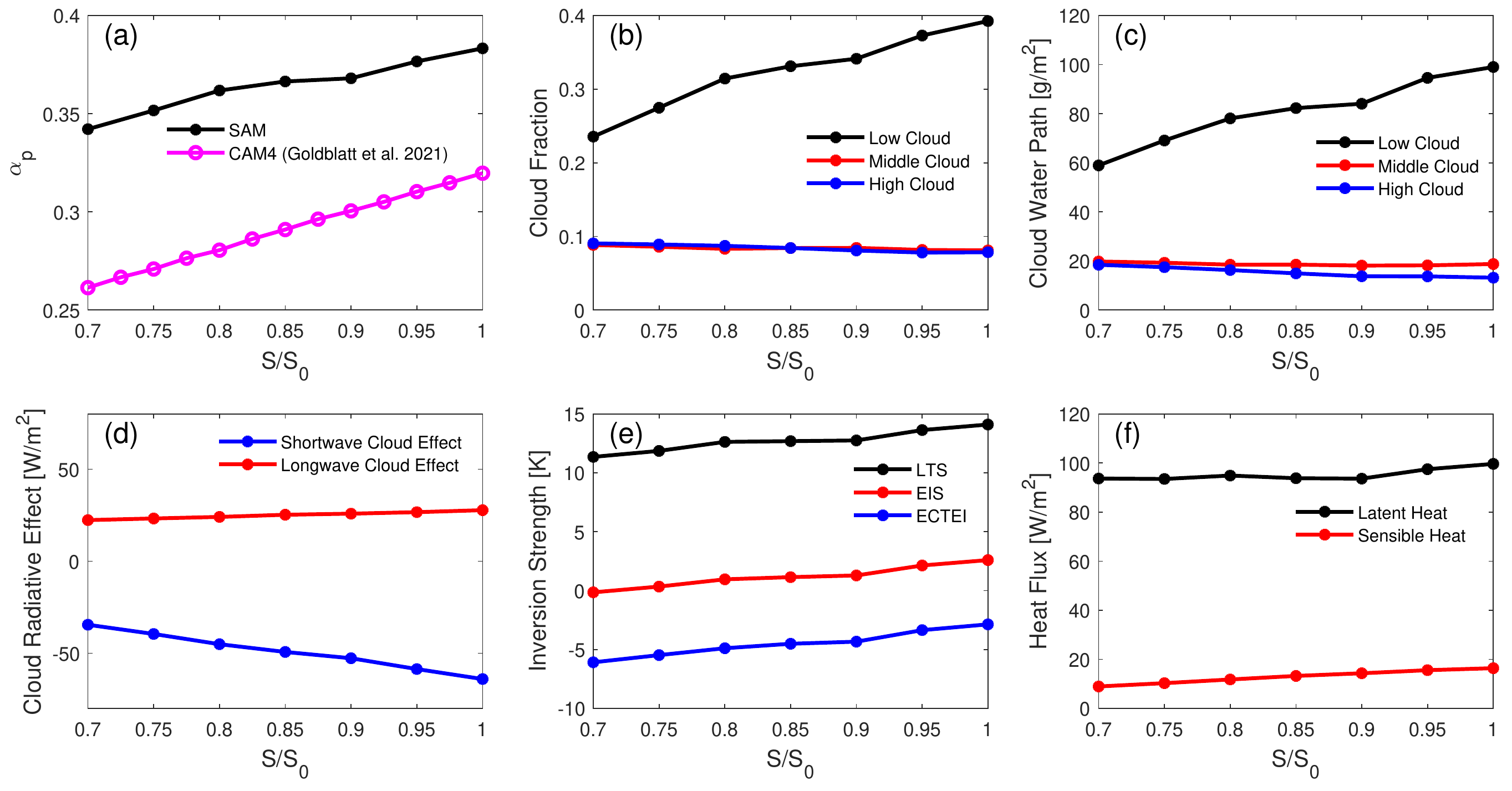}
\caption{Results under different solar radiations and corresponding CO$_2$ concentrations. (a) Global-mean planetary albedo in our SAM simulations (black), (b) cloud fractions of low-level (black), middle-level (red), and high-level clouds (blue), (c) same as (b) but for cloud water paths, (d) shortwave (blue) and longwave (red) cloud radiative effects, (e) inversion strength indexes: lower tropospheric stability (LTS; black), estimated inversion strength (EIS; red), and estimated cloud-top entrainment index (ECTEI; blue; see Text S3 in the Supporting Information online), and (f) surface latent (black) and sensible (red) heat fluxes. The low-level cloud fraction in (b) is defined as the fraction of grids with vertically-integrated cloud water path between surface and 700~hPa exceeding 0.02 kg m$^-{}^2$ \cite{Khairoutdinov2003}. Likewise, the middle-level and high-level cloud fractions correspond to the vertically-integrated cloud water path of 700 to 400~hPa and above 400~hPa. In (a), the CAM4 simulation results of \citeA{Goldblatt2021} (magenta) are included for comparisons.}
\label{globalmean}
\end{figure}

\subsection{Cloud Feedback}

\begin{figure}[t]
\noindent\includegraphics[width=\textwidth]{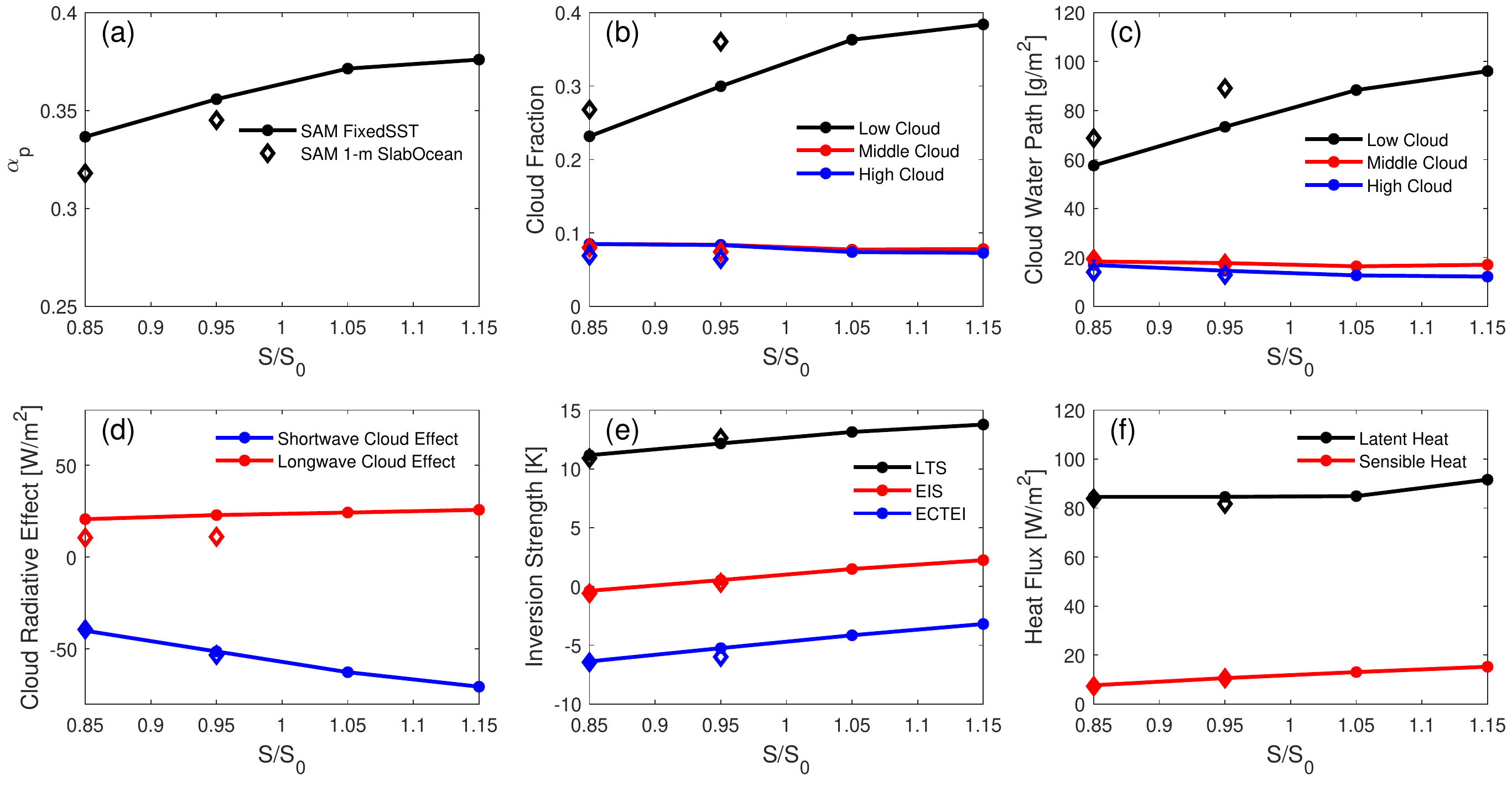}
\caption{Same as Figure~\ref{globalmean} but for the experiments in which the TOA and surface are nearly energy-balanced. The dot markers represent fixed-SST experiments, and the diamond markers represent slab ocean experiments.}
\label{globalmean_new}
\end{figure}

When insolation increases and meanwhile CO$_2$ concentration decreases, global-mean low-level cloud fraction significantly increases, high-level cloud fraction slightly decreases, and middle-level cloud fraction is almost unchanged (Figure~\ref{globalmean}b). The water paths of clouds at different levels show the same trends as the cloud fractions (Figure~\ref{globalmean}c). Moreover, the same trends of cloud fractions and cloud water paths can be found in the energy-balanced fixed-SST simulations and also in the slab ocean simulations (Figure~\ref{globalmean_new}).

The increase of low-level clouds can reflect more insolation to space, leading to a larger planetary albedo (Figure~\ref{globalmean}a). For example, comparing the 0.80S$_0$ case and the 1.00S$_0$ case, the planetary albedo increases from 0.36 to 0.38, and the magnitude of shortwave cloud radiative effect increases from 45.1 to 64.0 W\,m$^-{}^2$ (Figure~\ref{globalmean}d). The change of the shortwave cloud radiative effect is contributed by two parts, one is from the change of the solar radiation without any change of the cloud properties, and the other one is from the change of the cloud properties without any change of solar radiation. The albedo of the 1.00S$_0$ case is 0.38, so there should be $\approx$43.6 ($=1361.3/3.87\times(1.00-0.38)\times(1.0-0.8)$) W\,m$^-{}^2$ less solar radiation absorbed when the solar constant is 80\% of the modern Earth. Note that the denominator is 3.87 rather than 4.0; this is because the model uses Cartesian geometry rather than spherical geometry and meanwhile the polar regions are not simulated. When the 1.00S$_0$ case is chosen as the baseline, the shortwave cloud feedback is $-$45.1$-$($-$64.0)$\times$0.8/1.0 = 6.1 W\,m$^-{}^2$. In percentage, it is 14.0\% ($=$\,6.1/43.6). Another method for calculating the shortwave cloud feedback is shown in Text S2, and its value is 6.0~W\,m$^-{}^2$. For a solar constant of 70\% of the modern value, the strength of the shortwave cloud feedback is 10.3 W\,m$^-{}^2$ or 15.7\%. The magnitude of the shortwave cloud feedback is similar to \citeA{wolf2013}, in which it can contribute $\approx$9.6~W\,m$^-{}^2$ or 21\% when the Sun is 20\% dimmer (Figure~\ref{feedback}, \citeA{charnay2020}).

In the CAM4 simulations of \citeA{Goldblatt2021}, the shortwave cloud radiative effects are $-$33.4 and $-$52.5 W\,m$^-{}^2$, and the planetary albedos are 0.28 and 0.32, for the 0.80S$_0$ and 1.00S$_0$ cases, respectively (Figure S5). There should be 46.5 $(=1367.0/4.0\times(1.00-0.32)\times(1.0-0.8))$ W\,m$^-{}^2$ less solar radiation absorbed in the 0.80S$_0$ case. The shortwave cloud feedback equals $-$33.4$-$($-$52.5)$\times$0.8/1.0 = 8.6 W\,m$^-{}^2$, or 18.5\% ($=$\,8.6/46.5) in percentage. For 0.70S$_0$, the strength of the shortwave cloud feedback is 10.7 W\,m$^-{}^2$ or 15.4\% (Figure~\ref{feedback}). Note that the estimated magnitudes of the shortwave cloud feedback in \citeA{Goldblatt2021} are higher than those shown here, because the effect of varying solar constant on the value of shortwave cloud radiative effect without the changes of clouds was not excluded.

Moreover, the temperature patterns and the trends of cloud-related variables are in line with the CAM4 simulations in \citeA{Goldblatt2021} (Figures~S5 and S6). However, there are still some differences. For example, the change of high-level cloud fraction in CAM4 is larger than that in SAM (Figures~\ref{globalmean}b and S5b). Besides, cloud water paths in our simulations are lower than that in CAM4. The high-level cloud water path in SAM decreases from 0.70S$_0$ to 1.00S$_0$, but CAM4 shows an opposite trend (Figures~\ref{globalmean}c and S5c). The planetary albedo here is larger (Figure~\ref{globalmean}a), and the magnitude of the shortwave cloud radiative effect is also larger (Figures~\ref{globalmean}d and S5d).

Changing insolation and CO$_2$ concentration synchronously can also lead to changes in atmospheric temperature, atmospheric circulation, and cloud pattern. When we increase the insolation and meanwhile decrease the CO$_2$ concentration, the lower atmosphere becomes cooler but the upper atmosphere becomes warmer (Figure~\ref{climatology}d), the subsidence branch of the Hadley circulation becomes weaker (Figure~\ref{climatology}e), and the low-level clouds increase at almost every latitude (Figure~\ref{climatology}f). These patterns are similar to that in CAM4 shown in \citeA{Goldblatt2021}. But, the patterns in our simulations are more symmetric between the northern and southern hemispheres due to the aquaplanet we use.

\subsection{Mechanisms for Low-level Cloud Feedback}

Increasing insolation but meanwhile decreasing CO$_2$ concentration leads to the change of radiative heating rate, which can cause the change of clouds through several processes. Importantly, the lower troposphere experiences radiative cooling, but the upper troposphere experiences radiative warming (Figures~\ref{climatology}h and \ref{climatology}i), and subsequently the lower troposphere becomes cooler but the upper troposphere becomes warmer (Figures~\ref{climatology}d and \ref{climatology}g). To examine what factors dominate the changes in the radiative heating rates, we use an off-line radiative transfer model, RRTMG \cite{mlawer1997,clough2005}, to calculate the clear-sky longwave and shortwave heating rates under different conditions (Figure~S7). We find that decreasing CO$_2$ concentration leads to lower-tropospheric longwave radiative cooling, but the upper-tropospheric longwave heating rate doesn't change significantly (Figure~S7b). In our simulations, the changes of water vapor concentration are small, as well as its radiative effects (Figures~S7a, d, and e). For shortwave radiative heating rate, the effect of changing solar constant is greater than that of changing CO$_2$ or H$_2$O. When the solar radiation is increased, the clear-sky shortwave heating rate increases at all levels (Figure~S7f). In short, from 0.70S$_0$ to 1.00S$_0$, the enhanced lower-tropospheric radiative cooling dominated by less CO$_2$ and upper-tropospheric radiative warming dominated by more insolation make the lower troposphere and the upper troposphere become cooler and warmer, respectively.

Why there are more low-level clouds from 0.70S$_0$ to 1.00S$_0$? Following the analyzing methods of previous work (e.g., \citeA{bretherton2013, bretherton2015, qu2015, klein2017, mccoy2017,mieslinger2019}), we find four different processes, including boundary layer inversion, moisture gradient across the inversion, large-scale atmospheric circulation, and sensible heat flux and Bowen ratio. The details are as follows:
\begin{figure}[t]
\centering
\noindent\includegraphics[width=0.6\textwidth]{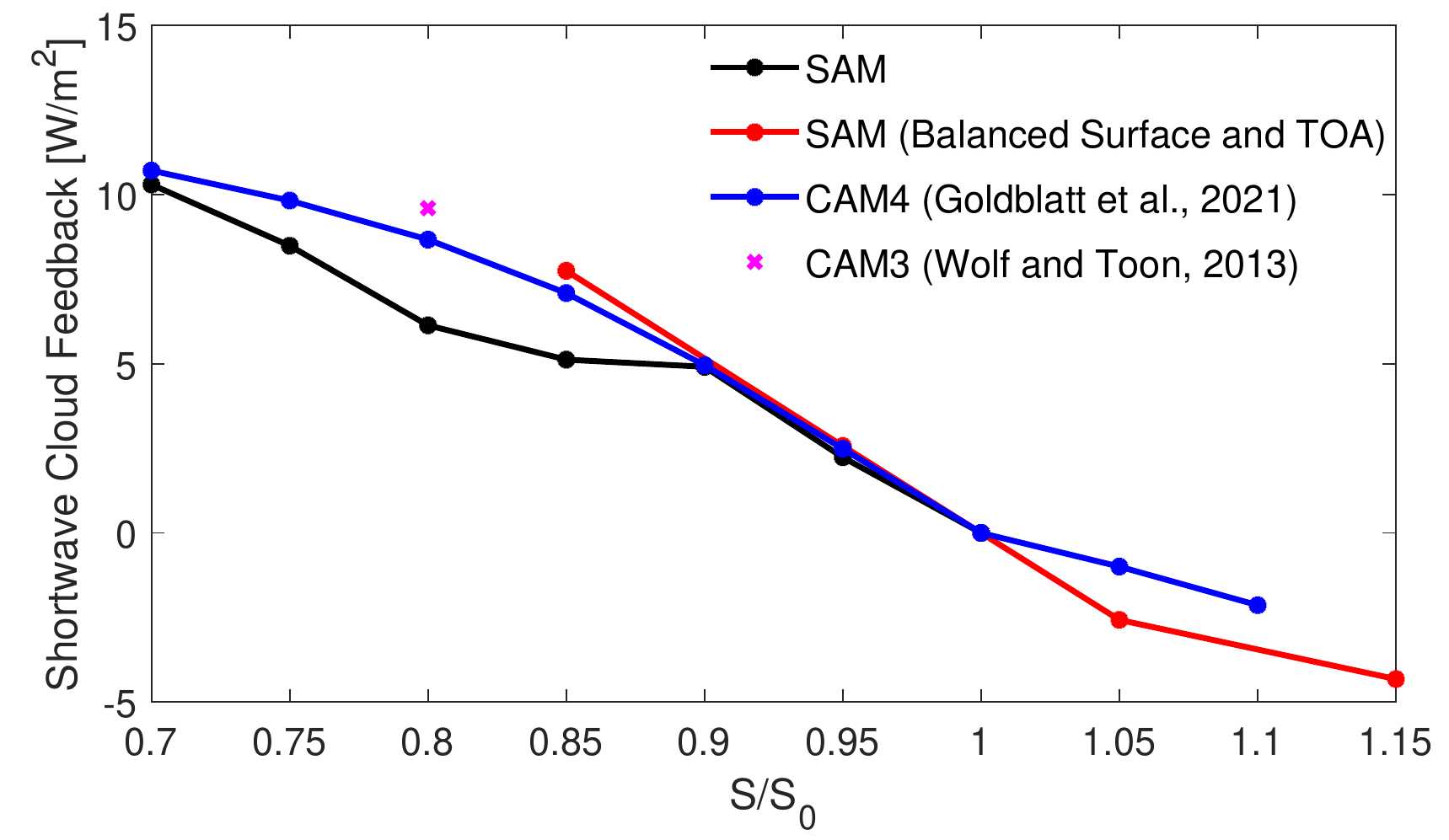}
\caption{The shortwave cloud feedback in different simulations. The 1.00S$_0$ case is chosen as the baseline, so its shortwave cloud feedback is zero. The magenta cross is the data from \citeA{wolf2013}, which used the general circulation model CAM3 but with a modified radiative transfer module.}
\label{feedback}
\end{figure}

\emph{Boundary layer inversion.} The lower troposphere becomes cooler while the upper troposphere becomes warmer from 0.70S$_0$ to 1.00S$_0$, so that the strength of the boundary layer inversion increases (Figure~\ref{climatology}g). These can also be seen from stability indicators, such as lower tropospheric stability (LTS), estimated inversion strength (EIS), and estimated cloud-top entrainment index (ECTEI) (\citeA{slingo1987,wood2006,kawai2017}; Text S3). All these three indicators show increased inversion strengths at all latitudes (Figures~\ref{globalmean}e and S8b-d). The stronger inversions can reduce dry air entrainment between the free troposphere and boundary layer and permit more boundary layer clouds \cite{klein1993,wood2006}.

\begin{figure}[t]
\includegraphics[width=\textwidth]{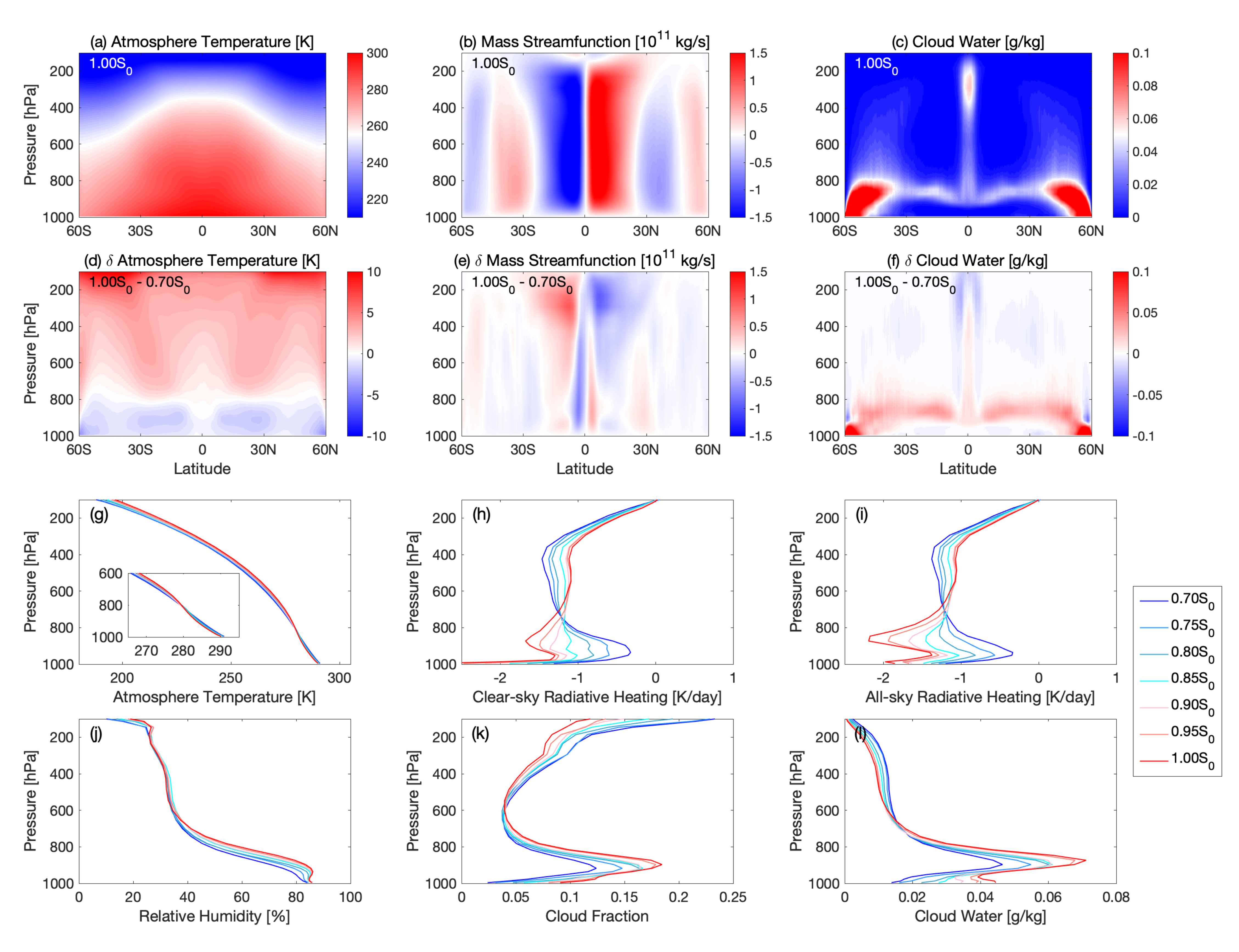}
\caption{Upper two rows: zonal-mean spatial patterns in the 1.00S$_0$ experiment (a--c) and the differences between 1.00S$_0$ and 0.70S$_0$ (d--f). From left to right, the variables are: atmospheric temperature, meridional mass streamfunction, and cloud water. Lower two rows: global-mean profiles of (g) atmosphere temperature, (h)--(i) clear-sky and all-sky radiative heating rate (shortwave plus longwave), (j) relative humidity, (k) cloud fraction, and (l) cloud water. Note that the cloud fraction in (k) is defined as the fraction of grids with cloud water larger than 0.01 g kg$^-{}^1$ \cite{wyant2009}.}
\label{climatology}
\end{figure}
\emph{Moisture gradient across the inversion.} The moisture gradient ($\delta q$) across the inversion is characterized by the difference in specific humidity between the surface and 700~hPa. Both air temperature and relative humidity (RH) at 700~hPa increase from 0.70S$_0$ to 1.00S$_0$, but near-surface temperature and RH nearly do not change (Figures~\ref{climatology}g and \ref{climatology}j), which leads to $\delta q$ gradually decreases (Figure~S8e). The smaller specific humidity gradient can lead to less dry entrainment across the inversion, which promotes low-level cloud maintenance \cite{van2015,scott2020,zhu2020}.

\emph{Large-scale atmospheric circulation.} The subsidence branch of the Hadley circulation becomes weaker from 0.70S$_0$ to 1.00S$_0$ (Figure~\ref{climatology}e). The more stable free troposphere (Figure~\ref{climatology}d) and the increase of static stability in the free troposphere may contribute to the weaker Hadley cells (e.g., \citeA{mitas2006,chemke2019}). The weaker subsidence allows the planetary boundary layer to extend to a higher altitude and thereby more low-level clouds can form \cite{bretherton2013,bretherton2015}.

\emph{Sensible heat flux and Bowen ratio.} Surface sensible heat flux slightly increases and latent heat flux keeps almost constant from 0.70S$_0$ to 1.00S$_0$ (Figures~\ref{globalmean}f and S8f). The larger temperature gradient near the surface layer (Figure~\ref{climatology}d) can contribute to larger sensible heat flux, but water vapor concentration near the surface layer which influences the latent heat flux nearly doesn't change (Figure~S7a). A larger sensible heat flux can increase the buoyancy and supply more heat from the surface to the boundary layer. Meanwhile, Bowen ratio, which is defined as the ratio of surface sensible heat flux to latent heat flux, increases from 0.70S$_0$ to 1.00S$_0$ (Figure~S9a). It can increase cloud water flux at cloud base, by influencing the efficiency of the heat cycle. Larger Bowen ratio corresponds to larger ratio of mechanical work to generate convection, supporting more low-level clouds (Figure~S9b, \citeA{sakradzija2017,mieslinger2019}).

The mechanisms addressed above can also be found in the four added fixed SST experiments and in the two slab ocean experiments (Figures~S10 and S11). This implies that our main conclusions are likely robust.

\section{Conclusions and Discussions}

In this study, we use a near-global cloud-permitting model to re-investigate the cloud feedback on Earth's long-term climate. We apply the grid spacing of 10~km $\times$ 14~km over a domain of half of the Earth's equatorial circumference in longitude and 60$^{\circ}$S to 60$^{\circ}$N in latitude. When solar radiation increases and meanwhile CO$_2$ concentration decreases, the change of low-level clouds dominates. The underlying mechanisms are analyzed from four different processes (summarized as Figure~S12): boundary layer inversion, moisture gradient across the inversion, large-scale atmospheric circulation, and surface heat flux and Bowen ratio. All these four factors contribute to the increase of low-level clouds.

The shortwave cloud feedback in our simulations can contribute $\approx$6~W\,m$^-{}^2$ or 14\% of the energy required to offset a fainter Sun that is 20\% dimmer than today. For a 30\% dimmer Sun, it can contribute $\approx$10~W\,m$^-{}^2$ or 16\%. This shortwave cloud feedback is likely to stabilize Earth's past or future long-term climate, and its impact is limited but unignorable.

Not only shortwave cloud radiative effect but also longwave cloud radiative effect changes in the experiments, but the magnitude of the former is much larger than the latter (Figure~\ref{globalmean}d). This is due to the fact that the response of low-level cloud water path is greater than that of the high-level cloud water path (Figures~\ref{globalmean}c, \ref{climatology}l, and S1d--f). Moreover, the concentration of CO$_2$ can also influence the strength of the longwave cloud radiative effect. When CO$_2$ concentration is higher, the longwave absorption spectrum is more saturated, which could lead to a weakening of the longwave cloud radiative effect even if there is no change in cloud properties \cite{wolf2013}. This is likely the reason why the cloud fraction and cloud water path of high-level clouds slightly decrease but the longwave cloud radiative effect slightly increases (Figures~2b--d). Unfortunately, the overlaps between CO$_2$ and cloud absorption spectra are not in model outputs. If these overlaps were considered, the longwave cloud feedback would be smaller than the change of the longwave cloud radiative effect shown in Figure~\ref{globalmean}d.

There is no continent distribution in the simulations, which would influence the location of ITCZ, the strength of the Hadley circulation, the surface heat fluxes, and other aspects. Future work should examine the role of land-sea distribution as well as use a dynamical ocean to test the effect of ocean dynamics. If continents are taken into account, other additional factors can affect the low-level cloud amount, such as zonal contrast of SST and horizontal energy transport \cite{norris1994,bretherton2013,myers2015}.

Accurately resolving the clouds needs a cloud-resolving model with a grid spacing of 4~km or less \cite{weisman1997}. Therefore, using the horizontal resolution of $\approx$10~km with no cumulus parameterization here may affect the strength of the cloud feedback. Besides, the conclusions of this study are based on a one-moment microphysics scheme, and different microphysics schemes may lead to different results, which needs to be examined in future work.

\acknowledgments
We thank Colin Goldblatt and the anonymous reviewer for their constructive comments. We thank Ji Nie, Yuwei Wang, and Jiachen Liu for their helpful discussions. Jun Yang is supported by the National Natural Science Foundation of China (NSFC) under grants 42075046 and 41888101.  

\noindent\textbf{Open Research}\\
The simulation data in this study are archived at \url{https://doi.org/10.5281/zenodo.6592041}.
The output data of CAM4 used for comparisons are included in \citeA{Goldblatt2021}, which can be accessed from \url{https://doi.org/10.20383/101.0308}.

\bibliography{cloudfeedback.bib}

\begin{thebibliography}{}

\bibitem [\protect \citeauthoryear {%
Bretherton%
}{%
Bretherton%
}{%
{\protect \APACyear {2015}}%
}]{%
bretherton2015}
\APACinsertmetastar {%
bretherton2015}%
\begin{APACrefauthors}%
Bretherton, C\BPBI S.%
\end{APACrefauthors}%
\unskip\
\newblock
\APACrefYearMonthDay{2015}{}{}.
\newblock
{\BBOQ}\APACrefatitle {Insights into low-latitude cloud feedbacks from
  high-resolution models} {Insights into low-latitude cloud feedbacks from
  high-resolution models}.{\BBCQ}
\newblock
\APACjournalVolNumPages{Philosophical Transactions of the Royal Society A:
  Mathematical, Physical and Engineering Sciences}{373}{2054}{20140415}.
\PrintBackRefs{\CurrentBib}

\bibitem [\protect \citeauthoryear {%
Bretherton%
, Blossey%
\BCBL {}\ \BBA {} Jones%
}{%
Bretherton%
\ \protect \BOthers {.}}{%
{\protect \APACyear {2013}}%
}]{%
bretherton2013}
\APACinsertmetastar {%
bretherton2013}%
\begin{APACrefauthors}%
Bretherton, C\BPBI S.%
, Blossey, P\BPBI N.%
\BCBL {}\ \BBA {} Jones, C\BPBI R.%
\end{APACrefauthors}%
\unskip\
\newblock
\APACrefYearMonthDay{2013}{}{}.
\newblock
{\BBOQ}\APACrefatitle {Mechanisms of marine low cloud sensitivity to idealized
  climate perturbations: {A} single-{LES} exploration extending the {CGILS}
  cases} {Mechanisms of marine low cloud sensitivity to idealized climate
  perturbations: {A} single-{LES} exploration extending the {CGILS}
  cases}.{\BBCQ}
\newblock
\APACjournalVolNumPages{Journal of Advances in Modeling Earth
  Systems}{5}{2}{316--337}.
\PrintBackRefs{\CurrentBib}

\bibitem [\protect \citeauthoryear {%
Bretherton%
\ \BBA {} Khairoutdinov%
}{%
Bretherton%
\ \BBA {} Khairoutdinov%
}{%
{\protect \APACyear {2015}}%
}]{%
bretherton2015b}
\APACinsertmetastar {%
bretherton2015b}%
\begin{APACrefauthors}%
Bretherton, C\BPBI S.%
\BCBT {}\ \BBA {} Khairoutdinov, M\BPBI F.%
\end{APACrefauthors}%
\unskip\
\newblock
\APACrefYearMonthDay{2015}{}{}.
\newblock
{\BBOQ}\APACrefatitle {Convective self-aggregation feedbacks in near-global
  cloud-resolving simulations of an aquaplanet} {Convective self-aggregation
  feedbacks in near-global cloud-resolving simulations of an
  aquaplanet}.{\BBCQ}
\newblock
\APACjournalVolNumPages{Journal of Advances in Modeling Earth
  Systems}{7}{4}{1765--1787}.
\PrintBackRefs{\CurrentBib}

\bibitem [\protect \citeauthoryear {%
Byrne%
\ \BBA {} Goldblatt%
}{%
Byrne%
\ \BBA {} Goldblatt%
}{%
{\protect \APACyear {2014}}%
}]{%
byrne2014}
\APACinsertmetastar {%
byrne2014}%
\begin{APACrefauthors}%
Byrne, B.%
\BCBT {}\ \BBA {} Goldblatt, C.%
\end{APACrefauthors}%
\unskip\
\newblock
\APACrefYearMonthDay{2014}{}{}.
\newblock
{\BBOQ}\APACrefatitle {Radiative forcings for 28 potential {Archean} greenhouse
  gases} {Radiative forcings for 28 potential {Archean} greenhouse
  gases}.{\BBCQ}
\newblock
\APACjournalVolNumPages{Climate of the Past}{10}{5}{1779--1801}.
\PrintBackRefs{\CurrentBib}

\bibitem [\protect \citeauthoryear {%
Charnay%
\ \protect \BOthers {.}}{%
Charnay%
\ \protect \BOthers {.}}{%
{\protect \APACyear {2013}}%
}]{%
charnay2013}
\APACinsertmetastar {%
charnay2013}%
\begin{APACrefauthors}%
Charnay, B.%
, Forget, F.%
, Wordsworth, R.%
, Leconte, J.%
, Millour, E.%
, Codron, F.%
\BCBL {}\ \BBA {} Spiga, A.%
\end{APACrefauthors}%
\unskip\
\newblock
\APACrefYearMonthDay{2013}{}{}.
\newblock
{\BBOQ}\APACrefatitle {Exploring the faint young {Sun} problem and the possible
  climates of the {Archean Earth} with a {3-D GCM}} {Exploring the faint young
  {Sun} problem and the possible climates of the {Archean Earth} with a {3-D
  GCM}}.{\BBCQ}
\newblock
\APACjournalVolNumPages{Journal of Geophysical Research:
  Atmospheres}{118}{18}{10--414}.
\PrintBackRefs{\CurrentBib}

\bibitem [\protect \citeauthoryear {%
Charnay%
, Wolf%
, Marty%
\BCBL {}\ \BBA {} Forget%
}{%
Charnay%
\ \protect \BOthers {.}}{%
{\protect \APACyear {2020}}%
}]{%
charnay2020}
\APACinsertmetastar {%
charnay2020}%
\begin{APACrefauthors}%
Charnay, B.%
, Wolf, E\BPBI T.%
, Marty, B.%
\BCBL {}\ \BBA {} Forget, F.%
\end{APACrefauthors}%
\unskip\
\newblock
\APACrefYearMonthDay{2020}{}{}.
\newblock
{\BBOQ}\APACrefatitle {Is the faint young {Sun} problem for {Earth} solved?}
  {Is the faint young {Sun} problem for {Earth} solved?}{\BBCQ}
\newblock
\APACjournalVolNumPages{Space Science Reviews}{216}{5}{1--29}.
\PrintBackRefs{\CurrentBib}

\bibitem [\protect \citeauthoryear {%
Chemke%
\ \BBA {} Polvani%
}{%
Chemke%
\ \BBA {} Polvani%
}{%
{\protect \APACyear {2019}}%
}]{%
chemke2019}
\APACinsertmetastar {%
chemke2019}%
\begin{APACrefauthors}%
Chemke, R.%
\BCBT {}\ \BBA {} Polvani, L\BPBI M.%
\end{APACrefauthors}%
\unskip\
\newblock
\APACrefYearMonthDay{2019}{}{}.
\newblock
{\BBOQ}\APACrefatitle {Opposite tropical circulation trends in climate models
  and in reanalyses} {Opposite tropical circulation trends in climate models
  and in reanalyses}.{\BBCQ}
\newblock
\APACjournalVolNumPages{Nature Geoscience}{12}{7}{528--532}.
\PrintBackRefs{\CurrentBib}

\bibitem [\protect \citeauthoryear {%
Clough%
\ \protect \BOthers {.}}{%
Clough%
\ \protect \BOthers {.}}{%
{\protect \APACyear {2005}}%
}]{%
clough2005}
\APACinsertmetastar {%
clough2005}%
\begin{APACrefauthors}%
Clough, S.%
, Shephard, M.%
, Mlawer, E.%
, Delamere, J.%
, Iacono, M.%
, Cady-Pereira, K.%
\BDBL {}Brown, P.%
\end{APACrefauthors}%
\unskip\
\newblock
\APACrefYearMonthDay{2005}{}{}.
\newblock
{\BBOQ}\APACrefatitle {Atmospheric radiative transfer modeling: {A} summary of
  the {AER} codes} {Atmospheric radiative transfer modeling: {A} summary of the
  {AER} codes}.{\BBCQ}
\newblock
\APACjournalVolNumPages{Journal of Quantitative Spectroscopy and Radiative
  Transfer}{91}{2}{233--244}.
\PrintBackRefs{\CurrentBib}

\bibitem [\protect \citeauthoryear {%
Collins%
\ \protect \BOthers {.}}{%
Collins%
\ \protect \BOthers {.}}{%
{\protect \APACyear {2006}}%
}]{%
collins2006}
\APACinsertmetastar {%
collins2006}%
\begin{APACrefauthors}%
Collins, W\BPBI D.%
, Rasch, P\BPBI J.%
, Boville, B\BPBI A.%
, Hack, J\BPBI J.%
, McCaa, J\BPBI R.%
, Williamson, D\BPBI L.%
\BDBL {}Zhang, M.%
\end{APACrefauthors}%
\unskip\
\newblock
\APACrefYearMonthDay{2006}{}{}.
\newblock
{\BBOQ}\APACrefatitle {The formulation and atmospheric simulation of the
  {Community Atmosphere Model version 3 (CAM3)}} {The formulation and
  atmospheric simulation of the {Community Atmosphere Model version 3
  (CAM3)}}.{\BBCQ}
\newblock
\APACjournalVolNumPages{Journal of Climate}{19}{11}{2144--2161}.
\PrintBackRefs{\CurrentBib}

\bibitem [\protect \citeauthoryear {%
Feulner%
}{%
Feulner%
}{%
{\protect \APACyear {2012}}%
}]{%
feulner2012}
\APACinsertmetastar {%
feulner2012}%
\begin{APACrefauthors}%
Feulner, G.%
\end{APACrefauthors}%
\unskip\
\newblock
\APACrefYearMonthDay{2012}{}{}.
\newblock
{\BBOQ}\APACrefatitle {The faint young {Sun} problem} {The faint young {Sun}
  problem}.{\BBCQ}
\newblock
\APACjournalVolNumPages{Reviews of Geophysics}{50}{2}{}.
\PrintBackRefs{\CurrentBib}

\bibitem [\protect \citeauthoryear {%
Goldblatt%
\ \protect \BOthers {.}}{%
Goldblatt%
\ \protect \BOthers {.}}{%
{\protect \APACyear {2009}}%
}]{%
goldblatt2009}
\APACinsertmetastar {%
goldblatt2009}%
\begin{APACrefauthors}%
Goldblatt, C.%
, Claire, M\BPBI W.%
, Lenton, T\BPBI M.%
, Matthews, A\BPBI J.%
, Watson, A\BPBI J.%
\BCBL {}\ \BBA {} Zahnle, K\BPBI J.%
\end{APACrefauthors}%
\unskip\
\newblock
\APACrefYearMonthDay{2009}{}{}.
\newblock
{\BBOQ}\APACrefatitle {Nitrogen-enhanced greenhouse warming on early {Earth}}
  {Nitrogen-enhanced greenhouse warming on early {Earth}}.{\BBCQ}
\newblock
\APACjournalVolNumPages{Nature Geoscience}{2}{12}{891--896}.
\PrintBackRefs{\CurrentBib}

\bibitem [\protect \citeauthoryear {%
Goldblatt%
, McDonald%
\BCBL {}\ \BBA {} E.%
}{%
Goldblatt%
\ \protect \BOthers {.}}{%
{\protect \APACyear {2021}}%
}]{%
Goldblatt2021}
\APACinsertmetastar {%
Goldblatt2021}%
\begin{APACrefauthors}%
Goldblatt, C.%
, McDonald, V\BPBI L.%
\BCBL {}\ \BBA {} E., M\BPBI K.%
\end{APACrefauthors}%
\unskip\
\newblock
\APACrefYearMonthDay{2021}{}{}.
\newblock
{\BBOQ}\APACrefatitle {Earth’s long-term climate stabilized by clouds}
  {Earth’s long-term climate stabilized by clouds}.{\BBCQ}
\newblock
\APACjournalVolNumPages{Nature Geoscience}{14}{}{143–150}.
\PrintBackRefs{\CurrentBib}

\bibitem [\protect \citeauthoryear {%
Goldblatt%
\ \BBA {} Zahnle%
}{%
Goldblatt%
\ \BBA {} Zahnle%
}{%
{\protect \APACyear {2011}}%
}]{%
goldblatt2011}
\APACinsertmetastar {%
goldblatt2011}%
\begin{APACrefauthors}%
Goldblatt, C.%
\BCBT {}\ \BBA {} Zahnle, K\BPBI J.%
\end{APACrefauthors}%
\unskip\
\newblock
\APACrefYearMonthDay{2011}{}{}.
\newblock
{\BBOQ}\APACrefatitle {Clouds and the {Faint Young Sun Paradox}} {Clouds and
  the {Faint Young Sun Paradox}}.{\BBCQ}
\newblock
\APACjournalVolNumPages{Climate of the Past}{7}{}{203-220}.
\PrintBackRefs{\CurrentBib}

\bibitem [\protect \citeauthoryear {%
Gough%
}{%
Gough%
}{%
{\protect \APACyear {1981}}%
}]{%
gough1981}
\APACinsertmetastar {%
gough1981}%
\begin{APACrefauthors}%
Gough, D\BPBI O.%
\end{APACrefauthors}%
\unskip\
\newblock
\APACrefYearMonthDay{1981}{}{}.
\newblock
{\BBOQ}\APACrefatitle {Solar interior structure and luminosity variations}
  {Solar interior structure and luminosity variations}.{\BBCQ}
\newblock
\APACjournalVolNumPages{Solar Physics}{74}{}{21-34}.
\PrintBackRefs{\CurrentBib}

\bibitem [\protect \citeauthoryear {%
Hartmann%
}{%
Hartmann%
}{%
{\protect \APACyear {2005}}%
}]{%
hartmann2005}
\APACinsertmetastar {%
hartmann2005}%
\begin{APACrefauthors}%
Hartmann, D\BPBI L.%
\end{APACrefauthors}%
\unskip\
\newblock
\APACrefYear{2005}.
\newblock
\APACrefbtitle {Global physical climatology} {Global physical climatology}.
\newblock
\APACaddressPublisher{}{Elsevier Science}.
\PrintBackRefs{\CurrentBib}

\bibitem [\protect \citeauthoryear {%
Heller%
, Duda%
, Winkler%
, Reitner%
\BCBL {}\ \BBA {} Gizon%
}{%
Heller%
\ \protect \BOthers {.}}{%
{\protect \APACyear {2021}}%
}]{%
heller2021}
\APACinsertmetastar {%
heller2021}%
\begin{APACrefauthors}%
Heller, R.%
, Duda, J\BHBI P.%
, Winkler, M.%
, Reitner, J.%
\BCBL {}\ \BBA {} Gizon, L.%
\end{APACrefauthors}%
\unskip\
\newblock
\APACrefYearMonthDay{2021}{}{}.
\newblock
{\BBOQ}\APACrefatitle {Habitability of the early {Earth}: liquid water under a
  faint young {Sun} facilitated by strong tidal heating due to a closer {Moon}}
  {Habitability of the early {Earth}: liquid water under a faint young {Sun}
  facilitated by strong tidal heating due to a closer {Moon}}.{\BBCQ}
\newblock
\APACjournalVolNumPages{PalZ}{95}{4}{563--575}.
\PrintBackRefs{\CurrentBib}

\bibitem [\protect \citeauthoryear {%
Kawai%
, Koshiro%
\BCBL {}\ \BBA {} Webb%
}{%
Kawai%
\ \protect \BOthers {.}}{%
{\protect \APACyear {2017}}%
}]{%
kawai2017}
\APACinsertmetastar {%
kawai2017}%
\begin{APACrefauthors}%
Kawai, H.%
, Koshiro, T.%
\BCBL {}\ \BBA {} Webb, M\BPBI J.%
\end{APACrefauthors}%
\unskip\
\newblock
\APACrefYearMonthDay{2017}{}{}.
\newblock
{\BBOQ}\APACrefatitle {Interpretation of factors controlling low cloud cover
  and low cloud feedback using a unified predictive index} {Interpretation of
  factors controlling low cloud cover and low cloud feedback using a unified
  predictive index}.{\BBCQ}
\newblock
\APACjournalVolNumPages{Journal of Climate}{30}{22}{9119--9131}.
\PrintBackRefs{\CurrentBib}

\bibitem [\protect \citeauthoryear {%
Khairoutdinov%
\ \BBA {} Kogan%
}{%
Khairoutdinov%
\ \BBA {} Kogan%
}{%
{\protect \APACyear {1999}}%
}]{%
khairoutdinov1999}
\APACinsertmetastar {%
khairoutdinov1999}%
\begin{APACrefauthors}%
Khairoutdinov, M\BPBI F.%
\BCBT {}\ \BBA {} Kogan, Y\BPBI L.%
\end{APACrefauthors}%
\unskip\
\newblock
\APACrefYearMonthDay{1999}{}{}.
\newblock
{\BBOQ}\APACrefatitle {A large eddy simulation model with explicit
  microphysics: Validation against aircraft observations of a
  stratocumulus-topped boundary layer} {A large eddy simulation model with
  explicit microphysics: Validation against aircraft observations of a
  stratocumulus-topped boundary layer}.{\BBCQ}
\newblock
\APACjournalVolNumPages{Journal of Atmospheric Sciences}{56}{13}{2115--2131}.
\PrintBackRefs{\CurrentBib}

\bibitem [\protect \citeauthoryear {%
Khairoutdinov%
\ \BBA {} Randall%
}{%
Khairoutdinov%
\ \BBA {} Randall%
}{%
{\protect \APACyear {2003}}%
}]{%
Khairoutdinov2003}
\APACinsertmetastar {%
Khairoutdinov2003}%
\begin{APACrefauthors}%
Khairoutdinov, M\BPBI F.%
\BCBT {}\ \BBA {} Randall, D\BPBI A.%
\end{APACrefauthors}%
\unskip\
\newblock
\APACrefYearMonthDay{2003}{}{}.
\newblock
{\BBOQ}\APACrefatitle {Cloud resolving modeling of the {ARM} summer 1997 {IOP}:
  Model formulation, results, uncertainties, and sensitivities} {Cloud
  resolving modeling of the {ARM} summer 1997 {IOP}: Model formulation,
  results, uncertainties, and sensitivities}.{\BBCQ}
\newblock
\APACjournalVolNumPages{Journal of the Atmospheric Sciences}{60}{}{607-625}.
\PrintBackRefs{\CurrentBib}

\bibitem [\protect \citeauthoryear {%
Klein%
, Hall%
, Norris%
\BCBL {}\ \BBA {} Pincus%
}{%
Klein%
\ \protect \BOthers {.}}{%
{\protect \APACyear {2017}}%
}]{%
klein2017}
\APACinsertmetastar {%
klein2017}%
\begin{APACrefauthors}%
Klein, S\BPBI A.%
, Hall, A.%
, Norris, J\BPBI R.%
\BCBL {}\ \BBA {} Pincus, R.%
\end{APACrefauthors}%
\unskip\
\newblock
\APACrefYearMonthDay{2017}{}{}.
\newblock
{\BBOQ}\APACrefatitle {Low-cloud feedbacks from cloud-controlling factors: A
  review} {Low-cloud feedbacks from cloud-controlling factors: A
  review}.{\BBCQ}
\newblock
\APACjournalVolNumPages{Surveys in Geophysics}{38}{}{1307--1329}.
\PrintBackRefs{\CurrentBib}

\bibitem [\protect \citeauthoryear {%
Klein%
\ \BBA {} Hartmann%
}{%
Klein%
\ \BBA {} Hartmann%
}{%
{\protect \APACyear {1993}}%
}]{%
klein1993}
\APACinsertmetastar {%
klein1993}%
\begin{APACrefauthors}%
Klein, S\BPBI A.%
\BCBT {}\ \BBA {} Hartmann, D\BPBI L.%
\end{APACrefauthors}%
\unskip\
\newblock
\APACrefYearMonthDay{1993}{}{}.
\newblock
{\BBOQ}\APACrefatitle {The seasonal cycle of low stratiform clouds} {The
  seasonal cycle of low stratiform clouds}.{\BBCQ}
\newblock
\APACjournalVolNumPages{Journal of Climate}{6}{8}{1587--1606}.
\PrintBackRefs{\CurrentBib}

\bibitem [\protect \citeauthoryear {%
Le~Hir%
, Teitler%
, Fluteau%
, Donnadieu%
\BCBL {}\ \BBA {} Philippot%
}{%
Le~Hir%
\ \protect \BOthers {.}}{%
{\protect \APACyear {2014}}%
}]{%
le2014}
\APACinsertmetastar {%
le2014}%
\begin{APACrefauthors}%
Le~Hir, G.%
, Teitler, Y.%
, Fluteau, F.%
, Donnadieu, Y.%
\BCBL {}\ \BBA {} Philippot, P.%
\end{APACrefauthors}%
\unskip\
\newblock
\APACrefYearMonthDay{2014}{}{}.
\newblock
{\BBOQ}\APACrefatitle {The faint young {Sun} problem revisited with a {3-D}
  climate--carbon model--{Part} 1} {The faint young {Sun} problem revisited
  with a {3-D} climate--carbon model--{Part} 1}.{\BBCQ}
\newblock
\APACjournalVolNumPages{Climate of the Past}{10}{2}{697--713}.
\PrintBackRefs{\CurrentBib}

\bibitem [\protect \citeauthoryear {%
McCoy%
, Eastman%
, Hartmann%
\BCBL {}\ \BBA {} Wood%
}{%
McCoy%
\ \protect \BOthers {.}}{%
{\protect \APACyear {2017}}%
}]{%
mccoy2017}
\APACinsertmetastar {%
mccoy2017}%
\begin{APACrefauthors}%
McCoy, D\BPBI T.%
, Eastman, R.%
, Hartmann, D\BPBI L.%
\BCBL {}\ \BBA {} Wood, R.%
\end{APACrefauthors}%
\unskip\
\newblock
\APACrefYearMonthDay{2017}{}{}.
\newblock
{\BBOQ}\APACrefatitle {The change in low cloud cover in a warmed climate
  inferred from {AIRS, MODIS, and ERA-Interim}} {The change in low cloud cover
  in a warmed climate inferred from {AIRS, MODIS, and ERA-Interim}}.{\BBCQ}
\newblock
\APACjournalVolNumPages{Journal of Climate}{30}{10}{3609--3620}.
\PrintBackRefs{\CurrentBib}

\bibitem [\protect \citeauthoryear {%
Mieslinger%
, Horv{\'a}th%
, Buehler%
\BCBL {}\ \BBA {} Sakradzija%
}{%
Mieslinger%
\ \protect \BOthers {.}}{%
{\protect \APACyear {2019}}%
}]{%
mieslinger2019}
\APACinsertmetastar {%
mieslinger2019}%
\begin{APACrefauthors}%
Mieslinger, T.%
, Horv{\'a}th, {\'A}.%
, Buehler, S\BPBI A.%
\BCBL {}\ \BBA {} Sakradzija, M.%
\end{APACrefauthors}%
\unskip\
\newblock
\APACrefYearMonthDay{2019}{}{}.
\newblock
{\BBOQ}\APACrefatitle {The dependence of shallow cumulus macrophysical
  properties on large-scale meteorology as observed in ASTER imagery} {The
  dependence of shallow cumulus macrophysical properties on large-scale
  meteorology as observed in aster imagery}.{\BBCQ}
\newblock
\APACjournalVolNumPages{Journal of Geophysical Research:
  Atmospheres}{124}{21}{11477--11505}.
\PrintBackRefs{\CurrentBib}

\bibitem [\protect \citeauthoryear {%
Mitas%
\ \BBA {} Clement%
}{%
Mitas%
\ \BBA {} Clement%
}{%
{\protect \APACyear {2006}}%
}]{%
mitas2006}
\APACinsertmetastar {%
mitas2006}%
\begin{APACrefauthors}%
Mitas, C\BPBI M.%
\BCBT {}\ \BBA {} Clement, A.%
\end{APACrefauthors}%
\unskip\
\newblock
\APACrefYearMonthDay{2006}{}{}.
\newblock
{\BBOQ}\APACrefatitle {Recent behavior of the {Hadley} cell and tropical
  thermodynamics in climate models and reanalyses} {Recent behavior of the
  {Hadley} cell and tropical thermodynamics in climate models and
  reanalyses}.{\BBCQ}
\newblock
\APACjournalVolNumPages{Geophysical Research Letters}{33}{1}{}.
\PrintBackRefs{\CurrentBib}

\bibitem [\protect \citeauthoryear {%
Mlawer%
, Taubman%
, Brown%
, Iacono%
\BCBL {}\ \BBA {} Clough%
}{%
Mlawer%
\ \protect \BOthers {.}}{%
{\protect \APACyear {1997}}%
}]{%
mlawer1997}
\APACinsertmetastar {%
mlawer1997}%
\begin{APACrefauthors}%
Mlawer, E\BPBI J.%
, Taubman, S\BPBI J.%
, Brown, P\BPBI D.%
, Iacono, M\BPBI J.%
\BCBL {}\ \BBA {} Clough, S\BPBI A.%
\end{APACrefauthors}%
\unskip\
\newblock
\APACrefYearMonthDay{1997}{}{}.
\newblock
{\BBOQ}\APACrefatitle {Radiative transfer for inhomogeneous atmospheres:
  {RRTM}, a validated correlated-k model for the longwave} {Radiative transfer
  for inhomogeneous atmospheres: {RRTM}, a validated correlated-k model for the
  longwave}.{\BBCQ}
\newblock
\APACjournalVolNumPages{Journal of Geophysical Research:
  Atmospheres}{102}{D14}{16663--16682}.
\PrintBackRefs{\CurrentBib}

\bibitem [\protect \citeauthoryear {%
Myers%
\ \BBA {} Norris%
}{%
Myers%
\ \BBA {} Norris%
}{%
{\protect \APACyear {2015}}%
}]{%
myers2015}
\APACinsertmetastar {%
myers2015}%
\begin{APACrefauthors}%
Myers, T\BPBI A.%
\BCBT {}\ \BBA {} Norris, J\BPBI R.%
\end{APACrefauthors}%
\unskip\
\newblock
\APACrefYearMonthDay{2015}{}{}.
\newblock
{\BBOQ}\APACrefatitle {On the relationships between subtropical clouds and
  meteorology in observations and {CMIP3} and {CMIP5} models} {On the
  relationships between subtropical clouds and meteorology in observations and
  {CMIP3} and {CMIP5} models}.{\BBCQ}
\newblock
\APACjournalVolNumPages{Journal of Climate}{28}{8}{2945--2967}.
\PrintBackRefs{\CurrentBib}

\bibitem [\protect \citeauthoryear {%
Norris%
\ \BBA {} Leovy%
}{%
Norris%
\ \BBA {} Leovy%
}{%
{\protect \APACyear {1994}}%
}]{%
norris1994}
\APACinsertmetastar {%
norris1994}%
\begin{APACrefauthors}%
Norris, J\BPBI R.%
\BCBT {}\ \BBA {} Leovy, C\BPBI B.%
\end{APACrefauthors}%
\unskip\
\newblock
\APACrefYearMonthDay{1994}{}{}.
\newblock
{\BBOQ}\APACrefatitle {Interannual variability in stratiform cloudiness and sea
  surface temperature} {Interannual variability in stratiform cloudiness and
  sea surface temperature}.{\BBCQ}
\newblock
\APACjournalVolNumPages{Journal of climate}{7}{12}{1915--1925}.
\PrintBackRefs{\CurrentBib}

\bibitem [\protect \citeauthoryear {%
Pierrehumbert%
}{%
Pierrehumbert%
}{%
{\protect \APACyear {2010}}%
}]{%
pierrehumbert2010}
\APACinsertmetastar {%
pierrehumbert2010}%
\begin{APACrefauthors}%
Pierrehumbert, R\BPBI T.%
\end{APACrefauthors}%
\unskip\
\newblock
\APACrefYear{2010}.
\newblock
\APACrefbtitle {Principles of planetary climate} {Principles of planetary
  climate}.
\newblock
\APACaddressPublisher{}{Cambridge University Press}.
\PrintBackRefs{\CurrentBib}

\bibitem [\protect \citeauthoryear {%
Qu%
, Hall%
, Klein%
\BCBL {}\ \BBA {} DeAngelis%
}{%
Qu%
\ \protect \BOthers {.}}{%
{\protect \APACyear {2015}}%
}]{%
qu2015}
\APACinsertmetastar {%
qu2015}%
\begin{APACrefauthors}%
Qu, X.%
, Hall, A.%
, Klein, S\BPBI A.%
\BCBL {}\ \BBA {} DeAngelis, A\BPBI M.%
\end{APACrefauthors}%
\unskip\
\newblock
\APACrefYearMonthDay{2015}{}{}.
\newblock
{\BBOQ}\APACrefatitle {Positive tropical marine low-cloud cover feedback
  inferred from cloud-controlling factors} {Positive tropical marine low-cloud
  cover feedback inferred from cloud-controlling factors}.{\BBCQ}
\newblock
\APACjournalVolNumPages{Geophysical Research Letters}{42}{18}{7767--7775}.
\PrintBackRefs{\CurrentBib}

\bibitem [\protect \citeauthoryear {%
Rosing%
, Bird%
, Sleep%
\BCBL {}\ \BBA {} Bjerrum%
}{%
Rosing%
\ \protect \BOthers {.}}{%
{\protect \APACyear {2010}}%
}]{%
rosing2010}
\APACinsertmetastar {%
rosing2010}%
\begin{APACrefauthors}%
Rosing, M\BPBI T.%
, Bird, D\BPBI K.%
, Sleep, N\BPBI H.%
\BCBL {}\ \BBA {} Bjerrum, C\BPBI J.%
\end{APACrefauthors}%
\unskip\
\newblock
\APACrefYearMonthDay{2010}{}{}.
\newblock
{\BBOQ}\APACrefatitle {No climate paradox under the faint early {Sun}} {No
  climate paradox under the faint early {Sun}}.{\BBCQ}
\newblock
\APACjournalVolNumPages{Nature}{464}{7289}{744--747}.
\PrintBackRefs{\CurrentBib}

\bibitem [\protect \citeauthoryear {%
Sagan%
\ \BBA {} Mullen%
}{%
Sagan%
\ \BBA {} Mullen%
}{%
{\protect \APACyear {1972}}%
}]{%
sagan1972}
\APACinsertmetastar {%
sagan1972}%
\begin{APACrefauthors}%
Sagan, C.%
\BCBT {}\ \BBA {} Mullen, G.%
\end{APACrefauthors}%
\unskip\
\newblock
\APACrefYearMonthDay{1972}{}{}.
\newblock
{\BBOQ}\APACrefatitle {Earth and {Mars}: evolution of atmospheres and surface
  temperatures} {Earth and {Mars}: evolution of atmospheres and surface
  temperatures}.{\BBCQ}
\newblock
\APACjournalVolNumPages{Science}{177}{}{52-56}.
\PrintBackRefs{\CurrentBib}

\bibitem [\protect \citeauthoryear {%
Sakradzija%
\ \BBA {} Hohenegger%
}{%
Sakradzija%
\ \BBA {} Hohenegger%
}{%
{\protect \APACyear {2017}}%
}]{%
sakradzija2017}
\APACinsertmetastar {%
sakradzija2017}%
\begin{APACrefauthors}%
Sakradzija, M.%
\BCBT {}\ \BBA {} Hohenegger, C.%
\end{APACrefauthors}%
\unskip\
\newblock
\APACrefYearMonthDay{2017}{}{}.
\newblock
{\BBOQ}\APACrefatitle {What determines the distribution of shallow convective
  mass flux through a cloud base?} {What determines the distribution of shallow
  convective mass flux through a cloud base?}{\BBCQ}
\newblock
\APACjournalVolNumPages{Journal of the Atmospheric
  Sciences}{74}{8}{2615--2632}.
\PrintBackRefs{\CurrentBib}

\bibitem [\protect \citeauthoryear {%
Scott%
\ \protect \BOthers {.}}{%
Scott%
\ \protect \BOthers {.}}{%
{\protect \APACyear {2020}}%
}]{%
scott2020}
\APACinsertmetastar {%
scott2020}%
\begin{APACrefauthors}%
Scott, R\BPBI C.%
, Myers, T\BPBI A.%
, Norris, J\BPBI R.%
, Zelinka, M\BPBI D.%
, Klein, S\BPBI A.%
, Sun, M.%
\BCBL {}\ \BBA {} Doelling, D\BPBI R.%
\end{APACrefauthors}%
\unskip\
\newblock
\APACrefYearMonthDay{2020}{}{}.
\newblock
{\BBOQ}\APACrefatitle {Observed sensitivity of low-cloud radiative effects to
  meteorological perturbations over the global oceans} {Observed sensitivity of
  low-cloud radiative effects to meteorological perturbations over the global
  oceans}.{\BBCQ}
\newblock
\APACjournalVolNumPages{Journal of Climate}{33}{18}{7717--7734}.
\PrintBackRefs{\CurrentBib}

\bibitem [\protect \citeauthoryear {%
Slingo%
}{%
Slingo%
}{%
{\protect \APACyear {1987}}%
}]{%
slingo1987}
\APACinsertmetastar {%
slingo1987}%
\begin{APACrefauthors}%
Slingo, J.%
\end{APACrefauthors}%
\unskip\
\newblock
\APACrefYearMonthDay{1987}{}{}.
\newblock
{\BBOQ}\APACrefatitle {The development and verification of a cloud prediction
  scheme for the {ECMWF} model} {The development and verification of a cloud
  prediction scheme for the {ECMWF} model}.{\BBCQ}
\newblock
\APACjournalVolNumPages{Quarterly Journal of the Royal Meteorological
  Society}{113}{477}{899--927}.
\PrintBackRefs{\CurrentBib}

\bibitem [\protect \citeauthoryear {%
Soden%
\ \BBA {} Held%
}{%
Soden%
\ \BBA {} Held%
}{%
{\protect \APACyear {2006}}%
}]{%
soden2006}
\APACinsertmetastar {%
soden2006}%
\begin{APACrefauthors}%
Soden, B\BPBI J.%
\BCBT {}\ \BBA {} Held, I\BPBI M.%
\end{APACrefauthors}%
\unskip\
\newblock
\APACrefYearMonthDay{2006}{}{}.
\newblock
{\BBOQ}\APACrefatitle {An assessment of climate feedbacks in coupled
  ocean–atmosphere models} {An assessment of climate feedbacks in coupled
  ocean–atmosphere models}.{\BBCQ}
\newblock
\APACjournalVolNumPages{Journal of Climate}{19}{}{3354–3360}.
\PrintBackRefs{\CurrentBib}

\bibitem [\protect \citeauthoryear {%
Van~der Dussen%
, De~Roode%
, Dal~Gesso%
\BCBL {}\ \BBA {} Siebesma%
}{%
Van~der Dussen%
\ \protect \BOthers {.}}{%
{\protect \APACyear {2015}}%
}]{%
van2015}
\APACinsertmetastar {%
van2015}%
\begin{APACrefauthors}%
Van~der Dussen, J.%
, De~Roode, S.%
, Dal~Gesso, S.%
\BCBL {}\ \BBA {} Siebesma, A.%
\end{APACrefauthors}%
\unskip\
\newblock
\APACrefYearMonthDay{2015}{}{}.
\newblock
{\BBOQ}\APACrefatitle {An {LES} model study of the influence of the free
  tropospheric thermodynamic conditions on the stratocumulus response to a
  climate perturbation} {An {LES} model study of the influence of the free
  tropospheric thermodynamic conditions on the stratocumulus response to a
  climate perturbation}.{\BBCQ}
\newblock
\APACjournalVolNumPages{Journal of Advances in Modeling Earth
  Systems}{7}{2}{670--691}.
\PrintBackRefs{\CurrentBib}

\bibitem [\protect \citeauthoryear {%
Vial%
, Bony%
, Stevens%
\BCBL {}\ \BBA {} Vogel%
}{%
Vial%
\ \protect \BOthers {.}}{%
{\protect \APACyear {2017}}%
}]{%
vial2017}
\APACinsertmetastar {%
vial2017}%
\begin{APACrefauthors}%
Vial, J.%
, Bony, S.%
, Stevens, B.%
\BCBL {}\ \BBA {} Vogel, R.%
\end{APACrefauthors}%
\unskip\
\newblock
\APACrefYearMonthDay{2017}{}{}.
\newblock
{\BBOQ}\APACrefatitle {Mechanisms and model diversity of trade-wind shallow
  cumulus cloud feedbacks: a review} {Mechanisms and model diversity of
  trade-wind shallow cumulus cloud feedbacks: a review}.{\BBCQ}
\newblock
\APACjournalVolNumPages{Surveys in Geophysics}{38}{6}{1331--1353}.
\PrintBackRefs{\CurrentBib}

\bibitem [\protect \citeauthoryear {%
Weisman%
\ \BBA {} Klemp%
}{%
Weisman%
\ \BBA {} Klemp%
}{%
{\protect \APACyear {1997}}%
}]{%
weisman1997}
\APACinsertmetastar {%
weisman1997}%
\begin{APACrefauthors}%
Weisman, W\BPBI C., M. L.and~Skamarock%
\BCBT {}\ \BBA {} Klemp, J\BPBI B.%
\end{APACrefauthors}%
\unskip\
\newblock
\APACrefYearMonthDay{1997}{}{}.
\newblock
{\BBOQ}\APACrefatitle {The resolution dependence of explicitly modeled
  convective systems} {The resolution dependence of explicitly modeled
  convective systems}.{\BBCQ}
\newblock
\APACjournalVolNumPages{Monthly Weather Review}{125}{}{527–548}.
\PrintBackRefs{\CurrentBib}

\bibitem [\protect \citeauthoryear {%
Wolf%
\ \BBA {} Toon%
}{%
Wolf%
\ \BBA {} Toon%
}{%
{\protect \APACyear {2013}}%
}]{%
wolf2013}
\APACinsertmetastar {%
wolf2013}%
\begin{APACrefauthors}%
Wolf, E.%
\BCBT {}\ \BBA {} Toon, O.%
\end{APACrefauthors}%
\unskip\
\newblock
\APACrefYearMonthDay{2013}{}{}.
\newblock
{\BBOQ}\APACrefatitle {Hospitable {A}rchean climates simulated by a general
  circulation model} {Hospitable {A}rchean climates simulated by a general
  circulation model}.{\BBCQ}
\newblock
\APACjournalVolNumPages{Astrobiology}{13}{7}{656--673}.
\PrintBackRefs{\CurrentBib}

\bibitem [\protect \citeauthoryear {%
Wood%
\ \BBA {} Bretherton%
}{%
Wood%
\ \BBA {} Bretherton%
}{%
{\protect \APACyear {2006}}%
}]{%
wood2006}
\APACinsertmetastar {%
wood2006}%
\begin{APACrefauthors}%
Wood, R.%
\BCBT {}\ \BBA {} Bretherton, C\BPBI S.%
\end{APACrefauthors}%
\unskip\
\newblock
\APACrefYearMonthDay{2006}{}{}.
\newblock
{\BBOQ}\APACrefatitle {On the relationship between stratiform low cloud cover
  and lower-tropospheric stability} {On the relationship between stratiform low
  cloud cover and lower-tropospheric stability}.{\BBCQ}
\newblock
\APACjournalVolNumPages{Journal of climate}{19}{24}{6425--6432}.
\PrintBackRefs{\CurrentBib}

\bibitem [\protect \citeauthoryear {%
Wordsworth%
\ \BBA {} Pierrehumbert%
}{%
Wordsworth%
\ \BBA {} Pierrehumbert%
}{%
{\protect \APACyear {2013}}%
}]{%
wordsworth2013}
\APACinsertmetastar {%
wordsworth2013}%
\begin{APACrefauthors}%
Wordsworth, R.%
\BCBT {}\ \BBA {} Pierrehumbert, R.%
\end{APACrefauthors}%
\unskip\
\newblock
\APACrefYearMonthDay{2013}{}{}.
\newblock
{\BBOQ}\APACrefatitle {Hydrogen-nitrogen greenhouse warming in {Earth's} early
  atmosphere} {Hydrogen-nitrogen greenhouse warming in {Earth's} early
  atmosphere}.{\BBCQ}
\newblock
\APACjournalVolNumPages{science}{339}{6115}{64--67}.
\PrintBackRefs{\CurrentBib}

\bibitem [\protect \citeauthoryear {%
Wyant%
, Bretherton%
\BCBL {}\ \BBA {} Blossey%
}{%
Wyant%
\ \protect \BOthers {.}}{%
{\protect \APACyear {2009}}%
}]{%
wyant2009}
\APACinsertmetastar {%
wyant2009}%
\begin{APACrefauthors}%
Wyant, M\BPBI C.%
, Bretherton, C\BPBI S.%
\BCBL {}\ \BBA {} Blossey, P\BPBI N.%
\end{APACrefauthors}%
\unskip\
\newblock
\APACrefYearMonthDay{2009}{}{}.
\newblock
{\BBOQ}\APACrefatitle {Subtropical low cloud response to a warmer climate in a
  superparameterized climate model. {Part I:} Regime sorting and physical
  mechanisms} {Subtropical low cloud response to a warmer climate in a
  superparameterized climate model. {Part I:} regime sorting and physical
  mechanisms}.{\BBCQ}
\newblock
\APACjournalVolNumPages{Journal of Advances in Modeling Earth Systems}{1}{3}{}.
\PrintBackRefs{\CurrentBib}

\bibitem [\protect \citeauthoryear {%
Zhu%
\ \BBA {} Poulsen%
}{%
Zhu%
\ \BBA {} Poulsen%
}{%
{\protect \APACyear {2020}}%
}]{%
zhu2020}
\APACinsertmetastar {%
zhu2020}%
\begin{APACrefauthors}%
Zhu, J.%
\BCBT {}\ \BBA {} Poulsen, C\BPBI J.%
\end{APACrefauthors}%
\unskip\
\newblock
\APACrefYearMonthDay{2020}{}{}.
\newblock
{\BBOQ}\APACrefatitle {On the increase of climate sensitivity and cloud
  feedback with warming in the {Community Atmosphere Models}} {On the increase
  of climate sensitivity and cloud feedback with warming in the {Community
  Atmosphere Models}}.{\BBCQ}
\newblock
\APACjournalVolNumPages{Geophysical Research Letters}{47}{18}{e2020GL089143}.
\PrintBackRefs{\CurrentBib}

\end{thebibliography}


\begin{thebibliography}{References From the Supporting Information}
    \bibitem[Goldblatt(2021)]{}Goldblatt, C., McDonald, V. L., \& E., M. K. (2021). Earth’s long-term climate stabilized by clouds. \emph{Nature Geoscience}, 14, 143–150.
    \bibitem[Kawai(2017)]{}Kawai, H., Koshiro, T., \& Webb, M. J. (2017). Interpretation of factors controlling low cloud cover and low cloud feedback using a unified predictive index. \emph{Journal of Climate}, 30 (22), 9119–9131.
    \bibitem[MacVean(1990)]{}MacVean, M., \& Mason, P. (1990). Cloud-top entrainment instability through small-scale mixing and its parameterization in numerical models. \emph{Journal of Atmospheric Sciences}, 47 (8), 1012–1030.
    \bibitem[Slingo(1987)]{}Slingo, J. (1987). The development and verification of a cloud prediction scheme for the ECMWF model. \emph{Quarterly Journal of the Royal Meteorological Society}, 113 (477), 899–927.
    \bibitem[Wood(2006)]{}Wood, R., \& Bretherton, C. S. (2006). On the relationship between stratiform low cloud cover and lower tropospheric stability. \emph{Journal of climate}, 19 (24), 6425–6432
    
\end{thebibliography}

\noindent\textbf{References From the Supporting Information}

\end{document}